\documentclass[%
reprint,
preprintnumbers,
 amsmath,amssymb,amsthm,superscriptaddress,
 aps,prd,
]{revtex4-2}

\usepackage{graphicx}% Include figure files
\usepackage{dcolumn}% Align table columns on decimal point
\usepackage{bm}% bold math
\usepackage{subcaption}
\usepackage{amsthm}
\usepackage[usenames]{color}
\definecolor{PineGreen}{rgb}{0.0,0.47,0.44}
\definecolor{MidnightBlue}{rgb}{0.1,0.1,0.44}
\definecolor{magenta}{rgb}{1.0,0.0,1.0}
\usepackage{listings}
\definecolor{org1}{rgb}{.92,.39,.21}
\definecolor{pur2}{rgb}{.53,.47,.7}
\usepackage{pgfplots}
\usepackage{wrapfig}
\usepackage{float}
\pgfplotsset{compat=newest}
\usepackage{hyperref}
\usepackage{caption}
\newcommand{\orcid}[1]{\href{https://orcid.org/#1}{\includegraphics[width=10pt]{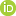}}}
\usepackage{footmisc}

\definecolor{myblue}{RGB}{72,127,227}
\definecolor{mygreen}{RGB}{48,142,48}
\definecolor{myviolet}{RGB}{57,49,103}
\definecolor{mymustard}{RGB}{213,147,62}

\definecolor{desycyan}{rgb}{0.00,0.68,0.93}
\definecolor{desyorange}{rgb}{0.96,0.52,0.07}
\definecolor{desygray}{rgb}{0.47,0.47,0.47}

\hypersetup{
	colorlinks=true,
	linkcolor=desycyan,
	citecolor=PineGreen,
	filecolor=magenta,
	urlcolor=desyorange
}

\newcommand{\RR}{\mathbb{R}}

\newcommand{\G}{\mathcal{G}}
\newcommand{\PP}{\mathbb{P}}

\newcommand{\CC}{\mathbb{C}}

\newcommand{\bV}{\mathbf{V}}

\newcommand{\cU}{\mathcal{U}}
\newcommand{\cF}{\mathcal{F}}
\newcommand{\cG}{\mathcal{G}}
\newcommand{\cI}{\mathcal{I}}

\newtheorem{theorem}{Theorem}

\newtheorem{definition}[theorem]{Definition}

\newtheoremstyle{example}{}{}{}{}{\bfseries}{\smallskip}{\newline}{}
\theoremstyle{example}

\usepackage{tikz-feynman}
\tikzfeynmanset{compat=1.1.0}

\newcommand{\hyperint}{\texttt{HyperInt} }
\newcommand{\xs}{.8}

\begin{document}
\preprint{DESY-24-023}

\title{Landau Singularities from Whitney Stratifications}%

\author{Martin Helmer\orcid{0000-0002-9170-8295}}
\thanks{\textcolor{black}{During the publication of this article, our dear friend and colleague Martin Helmer tragically passed away. We are grateful for many years of collaboration, and he will always be remembered.}}
\affiliation{Department of Mathematics, Swansea University, Swansea, Wales, UK}
\author{Georgios Papathanasiou\orcid{0000-0002-2627-9906}}
\email{g.papathanasiou@phys.uoa.gr}
\affiliation{Department of Physics, National and Kapodistrian University of Athens, Athens 15784, Greece}
\affiliation{Department of Mathematics, City, University of London,
Northampton Square, EC1V 0HB, London, UK}
\affiliation{Deutsches Elektronen-Synchrotron DESY, Notkestr.~85, 22607 Hamburg, Germany.}
\author{Felix Tellander\orcid{0000-0001-6418-8047}}
\email{felix@tellander.se}
\affiliation{Deutsches Elektronen-Synchrotron DESY, Notkestr.~85, 22607 Hamburg, Germany.}
\affiliation{Mathematical Institute, University of Oxford, Oxford OX2 6GG, UK}
\affiliation{School of Mathematics and Hamilton Mathematics Institute, Trinity College, Dublin 2, Ireland}


\begin{abstract}
We demonstrate that the complete and non-redundant set of Landau singularities of Feynman integrals may be explicitly obtained from the Whitney stratification of a certain map. As a proof of concept, we leverage recent theoretical and algorithmic advances in their computation in order to determine this set for nontrivial examples of two-loop integrals. Interestingly, different strata of the Whitney stratification describe not only the singularities of a given integral, but also those of integrals obtained from kinematic limits, e.g.~by setting some of its masses or momenta to zero.
\end{abstract}

\maketitle

%\tableofcontents
%%%%%%%%%%%%%%%%%%%%%%%%%%%%%%%%%%%%%%%%%%%%%%%%%%%%%%%%%%%%%%%%%%%%%%%%%%%%%%%%%%%%%%%%%%
%\section{\label{sec: introduction} Introduction}
At the heart of both cross-section calculations at the Large Hadron Collider and gravitational wave physics lie the evaluation of \emph{Feynman integrals}. These integrals are multivalued meromorphic functions of their kinematic variables, and understanding their analytic structure has been an ongoing quest for theoretical physicists since the late 50's \cite{Eden:1966dnq}.  

Key information on the analytic structure of Feynman integrals is provided by the values of the kinematic variables for which they become singular, first studied in the pioneering work of Landau~\cite{Landau:1959fi}. A virtue of these \emph{Landau singularities} is that knowing them in advance may significantly aid the evaluation of the integrals, for example with the canonical differential equations approach~\cite{Henn:2013pwa}: They may constrain or even fully predict~\cite{Dlapa:2023cvx} the analytic building blocks or (symbol) \emph{letters}~\cite{Goncharov:2010jf} of this approach, thereby turning the determination of the differential equations from a symbolic, to a much simpler numerical problem. Thanks to their utility, also in other aspects of Feynman integration~\cite{Gardi:2022khw}, Landau analysis is currently experiencing a revival with ever-increasing momentum, see for example~\cite{Brown:2009ta,Panzer:2014caa,Klausen:2021yrt,Hannesdottir:2022xki,Mizera:2021icv,Berghoff:2022mqu,Dlapa:2023cvx,Fevola:2023short,Fevola:2023long}.

A mathematically robust definition of what a Landau singularity is has been provided by Pham \cite{Pham2011integrals} as his \emph{Landau variety}; roughly speaking it is characterized by the critical values of a projection map from the variety describing the integrand in terms of  all integration and kinematic variables, to the space of just the kinematic variables.
Its direct calculation from this definition has proven quite challenging, and much of the recent effort has focused on developing alternative methods that more simply compute varieties that either contain it~\cite{Brown:2009ta,Panzer:2014caa}, or are contained in it~\cite{Dlapa:2023cvx,Fevola:2023short,Fevola:2023long}. In other words, these methods may in general either provide spurious candidates or miss certain Landau singularities entirely, and it is not known when these phenomena do not occur beyond certain special classes of integrals.

In this work, we demonstrate that the rigorous definition of Landau singularities \emph{can} be used to obtain their complete set, and nothing but their complete set, in a practical manner. To this end, we apply recent advances on the computation of Whitney stratifications \cite{Helmer2021conormal,helmer2023effective}, which enter this definition. In particular, we leverage their implementation in the \texttt{WhitneyStratifications} 2.27 package for the  Macaulay2~\cite{M2} software, available at \url{http://martin-helmer.com/Software/WhitStrat/index.html}, in order to compute the Landau singularities of nontrivial two-loop integrals, such as the two-mass hard slashed box and the parachute. While this computational route is currently not as efficient as that of other specialized software for sub- or supersets of Landau singularities~\cite{Panzer:2014caa,Fevola:2023short,Fevola:2023long}, it provides a proof of concept that paves the way for their fast and accurate determination in the future.

%%%%%%%%%%%%%%%%%%%%%%%%%%%%%%%%%%%%%%%%%%%%%%%%%%%%%%%%%%%%%%%%%%%%%%%%%%%%%%%%%%%%%%%%%

\begin{figure*}%
\captionsetup[subfigure]{justification=centering}
    \begin{subfigure}[t]{0.3\textwidth}
        \centering
         \includegraphics[width=0.6\linewidth]{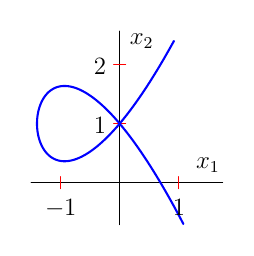}%
%             \captionsetup{justification=raggedright,
%singlelinecheck=false
%}
        \caption{$z=-1$: Curve with one loop.}
        \label{fig:curve node}
    \end{subfigure}%
    \begin{subfigure}[t]{0.3\textwidth}
        \centering
         \includegraphics[width=0.6\linewidth]{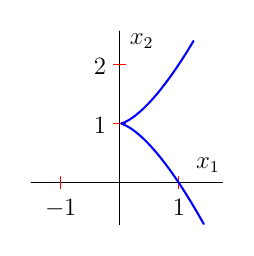}%
             %\captionsetup{justification=raggedright,
%singlelinecheck=false
%}
        \caption{$z=0$:  Curve with a cusp.}%
        \label{fig: curve cusp}
    \end{subfigure}%
    \begin{subfigure}[t]{0.3\textwidth}
        \centering
        \includegraphics[width=0.6\linewidth]{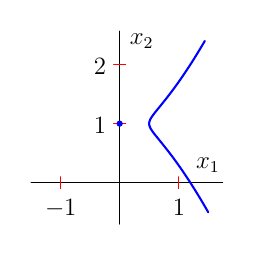}%
%                    \captionsetup{justification=raggedright,
%singlelinecheck=false
%}
    \caption{$z=1/2$: Smooth curve with two connected components. }
        \label{fig: curve smooth}
    \end{subfigure}%
                \captionsetup{justification=raggedright,
singlelinecheck=false
}
    \caption{Example of how Landau singularities of integrals are captured by the changes of their integrand's topology. Here, the integrand \eqref{eq:planarCubic} defines a curve in the integration variables $x_i$, whose topology changes when the external parameter takes the value $z=0$. This precisely coincides with the integral's Landau singularity.}
\label{fig:curve}\vspace{-4mm}
\end{figure*}

\section*{\label{sec: Feynman integrals} Feynman integrals and their singularities}
In this article we consider one-particle irreducible Feynman graphs $G:=(E,V)$ with set of edges and vertices $E$ and $V$, respectively,  and loop number $L=n-|V|+1$, where we abbreviate $n:=|E|$. The vertex set $V$ has the disjoint partition $V=V_{\mathrm{ext}}\sqcup V_{\mathrm{int}}$ where each vertex $v\in V_{\mathrm{ext}}$ is assigned an external incoming $4$-dimensional momentum $p_v\in\RR^{1,3}$, and each internal edge $e$ is assigned a mass parameter $m_e$. Using dimensional regularization with $D:=4-2\epsilon$ and the \emph{Lee-Pomeransky representation} \cite{Lee2013}, we assign the following integral to $G$:
\begin{equation}{\label{eq: Feynman integral}}    \cI=\int_{\RR_+^n}\left(\prod_{i=1}^n\frac{x_i^{\nu_i}\,dx_i}{x_i\Gamma(\nu_i)}\right)\,\frac{1}{\cG^{D/2}},\qquad \cG=\cU+\cF,
\end{equation}
where $\nu_i$ are propagator powers and $\cU$, respectively $\cF$, are the first and second \emph{Symanzik polynomials}. These 
are homogeneous polynomials of degree $L$ and $L+1$ in the $x_i$, respectively, and may be easily obtained from graph-theoretic data, see e.g.~\cite{Weinzierl:2022eaz}. The polynomial coefficients of $\cU$ are just numbers, whereas those of $\cF$ depend on the kinematic variables $p_v, m_e$, and are not necessarily algebraically independent.

We will render the representation~\eqref{eq: Feynman integral} of Feynman integrals projective, similarly to how this is done in the Feynman parameter representation~\cite{Panzer:2015ida}, see also~\cite{Chen:2024xwt}. Namely, we  insert $1=\int x_0^{\nu_0}\delta(1-x_0)dx_0$ in \eqref{eq: Feynman integral} and, dropping the gamma functions, we express it as an integral of the \emph{homogenized} Lee-Pomeransky polynomial $\cG_h$,
\begin{equation}\label{eq: projective Feynman integral}
    \cI=\int_{\PP_+^n}\frac{\prod_{i=0}^nx_i^{\nu_i}}{\cG_h^{D/2}}\,\Omega\,,\quad \cG_h:=\cU x_0+\cF\,,
\end{equation}
over the \emph{projective simplex} $\PP_+^n=\{[x_0:x_1:\cdots:x_n]\in\RR\PP^n\,|\,x_i>0,\ i=0,\,1,\,\ldots,\ n]\}$. In this equation, $\Omega=\sum_{i=0}^n(-1)^{n-1-i}dx_0/x_0\wedge\cdots\wedge\widehat{dx_i}/x_i\wedge\cdots\wedge dx_n/x_n$, where the hatted term is omitted, and $\nu_0=(L+1)D/2-\sum_{i=1}^n\nu_i,$ ensuring projective invariance.

Singularities of Feynman integrals were initially formulated in terms of conditions for their contour of integration to become trapped between colliding poles of the integrand. These conditions are known as the Landau equations~\cite{Landau:1959fi}, which in the representation \eqref{eq: projective Feynman integral} are,
\begin{equation}\label{eq: Landau condition}
    \cG_h=0\ \mathrm{and}\  x_i\frac{\partial\cG_h}{\partial x_i}=0\ \forall\ i=0,\,1,\,\ldots,\,n.
\end{equation}
Due to their generically nonlinear nature, directly and efficiently solving the above equations remains a challenge. When the polynomials $\cU$ and $\cF$ are taken to have generic coefficients, then Landau singularities are alternatively captured by the \emph{principal} $A$\emph{-determinant}~\cite{gelfand2008discriminants}, see also~\cite{Klausen:2021yrt}. This is a polynomial in the kinematic variables, which vanishes whenever the equations~\eqref{eq: Landau condition} have a solution.

While the algorithmic nature of the principal $A$-determinant is appealing, an issue arises when specialising the polynomial coefficients back to their Feynman integral values: in this process certain factors of the principal $A$-determinant may vanish identically.  Given the resulting unphysical conclusion, i.e.~that the Feynman integral is singular for all values of the kinematic variables, one may be tempted to define its Landau singularities by simply removing these vanishing factors. However this definition is incomplete, as it has been definitively shown to miss singularities~\cite{Berghoff:2022mqu}. An empirical refinement of this definition, based on sequential limits of the generic polynomial coefficients, has also been proposed and  tested in many one- and some two-loop integrals~\cite{Dlapa:2023cvx}, however a more extensive vetting and/or a proof would be desirable. Another closely related refinement, known as the \emph{principal Landau determinant}, has also been recently defined and implemented in the Julia package \texttt{PLD.jl}~\cite{Fevola:2023short,Fevola:2023long}; unfortunately, this also misses singularities in certain cases.

A different approach to singularities of integrals, due to Pham, boils down to considering the subspace of integration variables where the integrand denominator vanishes, and detecting for which values of the external parameters this variety changes topology. We illustrate the essence of this approach with the following simple example of an the integrand denominator,
\begin{equation}
   f_z(x_1,x_2)=(x_2-1)^2-(x_1-z)x_1^2\,, \label{eq:planarCubic}
\end{equation}
where $x_1,x_2$ are the integration variables and $z$ the external (kinematic) parameter.
%~\footnote{This example was also previously used to demonstrate the incompleteness of the aforementioned principal Landau discriminant approach.}
The variety  defined by $f_z(x_1,x_2)=0$ is the curve shown in Figure \ref{fig:curve} for different values of $z$. Clearly, the topology of the curve changes at $z=0$, and this is precisely where the integral develops a Landau singularity.

In Pham's setting~\cite{Pham2011integrals}, one may equivalently consider the same variety but in the space of both integration variables and parameters, and examine the map projecting it to the space of parameters. The critical values where this projection map drops rank then define the Landau variety, capturing the aforementioned changes in topology. In the example \eqref{eq:planarCubic}, we apply the projection map $(x_1,x_2,z)\mapsto z$ from the surface $f_z(x_1,x_2)=0$ inside the $x_1,x_2,z$ space, this map's critical value is exactly $z=0$.

Our work is motivated by the fact that no general recipe existed for directly computing the Landau variety. An upper bound for the latter is provided by Brown's polynomial reduction~\cite{Brown:2009ta}, also implemented in \hyperint~\cite{Panzer:2014caa}. It has subsequently been proposed that this superset may be reduced to the Landau variety by only keeping the components where the \emph{Euler characteristic} of the parametric variety describing the integrand changes from its generic value~\cite{Fevola:2023short,Fevola:2023long}. Closely related to the number of handles of a surface, the Euler characteristic $\chi$ is a simple sufficient measure of changes in its topology, and the aforementioned proposal may be proven correct as a consequence  of either \cite{esterov2013discriminant} or \cite{AMENDOLA2019222} in the special case where the Landau variety is equal to the zero set of the principal $A$-determinant. More generally, however, it is not currently known if this proposal fails in cases where a change in topology leaves the Euler characteristic invariant.

Here, we will instead show that the Landau variety may be computed directly and completely algorithmically, by means of the {Whitney stratification} of the projection map discussed two paragraphs above. This is summarised in Definition~\ref{def: landau variety}, and in what follows we introduce the necessary foundations.

%%%%%%%%%%%%%%%%%%%%%%%%%%%%%%%%%%%%%%%%%%%%%%%%%%%%%%%%%%%%%%

\section*{From Whitney to Landau}
Let $\mathbb{K}$ be either the  field of real or complex numbers. In the discussion that follows it will be convenient to employ the following notation for the {\em algebraic variety} defined by polynomials $g_1, \dots, g_r$ in a polynomial ring in the variables $x=(x_1,\ldots,x_n)$ over $\mathbb{K}$, $\mathbb{K}[x_1,\dots, x_n]$: 
$$ 
\bV(g_1, \dots, g_r):= \{ x\in \mathbb{K}^n\; |\; g_1(x)=\cdots = g_r(x)=0\}.
$$
We will now proceed to define the Whitney stratification of a potentially singular variety, which loosely speaking decomposes it into smooth components such that the local geometry is the same for any two points of the same connected component.
Consider an algebraic variety $X$ of dimension $d$. Then a flag $X_\bullet$ of varieties $X_0\subset \cdots \subset X_d=X$ contained in one another is a {\em Whitney stratification} of $X$ if the (open) sets $X_i-X_{i-1}$
, consisting of all points in $X_i$ which are not in $X_{i-1}$
 are smooth manifolds for all $i$, and {\em Whitney's condition} B holds for all pairs of connected components or \emph{strata} of these manifolds. A pair of strata, $M,N$, whose closures obey $\overline{M}\subset \overline{N}$, satisfy Whitney's condition B at a point $x\in M$ with respect to $N$ if for every sequence $\{p_\ell\} \subset M$ and $\{q_\ell\}\subset N$ with $\lim p_\ell=\lim q_\ell=x$, 
  the limit of secant lines between $p_\ell,q_\ell$ is contained in the limit of tangent planes to $N$ at $q_\ell$. Furthermore, the pair $M, N$ satisfies condition B if condition B holds, with respect to $N$, at all points $x\in M$.

\begin{figure}\centering
    \begin{tikzpicture}
    % Include the image
    \node[inner sep=0] (image) at (0,0) {\includegraphics[scale=0.18,trim={18cm 1cm 12cm 7cm},clip]{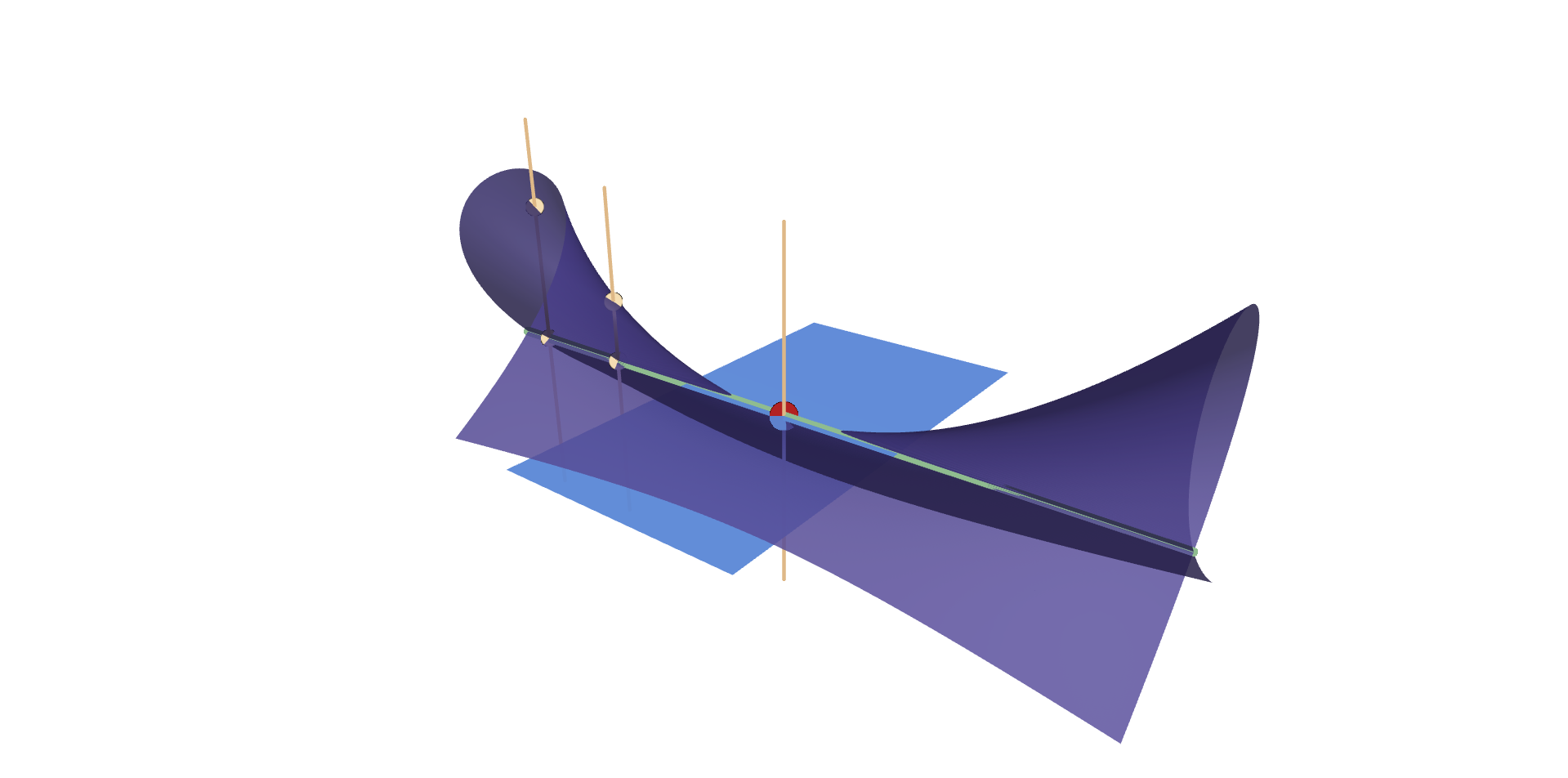}};

    % Draw the coordinate axes
    \begin{scope}[shift={(-3,-2)}] % Adjust the shift to place the axes correctly
        \draw[-stealth] (0,0,0) -- (0.86,-0.29,0) node[anchor=north east] {$y$};
        \draw[-stealth] (0,0,0) -- (0,1,0) node[anchor=north west] {$z$};
        \draw[-stealth] (0,0,0) -- (0,0.11,1.62) node[above=2pt] {$x$};
    \end{scope}
    \node[align=center] at (-3,-2) (ori) {};
\draw[dashed, gray] (ori) -- (-0.5,0.5,0);
\end{tikzpicture}

    \caption{The Whitney stratification of the {\em \textcolor{violet}{Whitney cusp}} $X=\bV(x^2 + z^3-y^2z^2)$  is  $X\supset \bV(x,z)\supset \{{(0,0,0)}\}$. The \textcolor{red}{origin} must be a separate stratum since the limiting \textcolor{myblue}{tangent plane} of a sequence approaching it from the top left does not contain the limiting \textcolor{mymustard}{secant line} between this and the sequence on the negative \textcolor{mygreen}{$y$-axis}.
    \label{fig:WhitCusp2}}
\end{figure}

A concrete example illustrating these definitions is the Whitney stratification of the \textcolor{violet}{Whitney cusp},
\begin{equation*}
X=\bV(x^2 + z^3-y^2z^2)\,,
\end{equation*}
 shown in Figure~\ref{fig:WhitCusp2}, where the color-coding used here and below matches that of the figure. Notice that while this surface would still be decomposed into smooth pieces if we took the entire \textcolor{mygreen}{$y$-axis} $\mathbf{V}(x,z)$ as a component. However, the local geometry at the \textcolor{red}{origin} $\{{(0,0,0)}\}=\mathbf{V}(x,y,z)$ is different, as the loop in the $(x,z)$-plane contracts to a point. Whitney's condition B captures this, as two sequences approaching the \textcolor{red}{origin}, one on the negative \textcolor{mygreen}{$y$-axis} and one on the component above it, have a limiting \textcolor{mymustard}{secant line} that does not lie on the limiting \textcolor{myblue}{tangent plane} of the latter component. Therefore the \textcolor{red}{origin} needs to be placed in a separate stratum. Together with the two half-infinite lines of the \textcolor{mygreen}{$y$-axis} and the four two-dimensional strata of the \textcolor{violet}{Whitney cusp},  the full Whitney stratification in the field of reals is thus
\begin{equation*}
    X\supset\bV(x,z)\supset\{{(0,0,0)}\}.
\end{equation*}

Note that a Whitney stratification of a variety is not unique, however it was shown in \cite{teissier1982varietes} that when $X$ is any complex algebraic (or analytic) variety there exists a {\bf unique} {\em minimal (or coarsest) Whitney stratification},  such that all other Whitney stratifications are obtained by adding strata inside the strata of the minimal one, see also \cite{FTpolar}.

%Although Whitney stratifications of varieties were used, at least implicitly, in the original definition of the Landau variety, it is the concept of a {\em Whitney stratification of an algebraic map} which will allow us to compute the latter in a completely systematic manner. We define this as follows: 
Next, we consider an algebraic map $f:X\to Y$ between varieties $X$ and $Y$. A  \emph{Whitney stratification of the map $f$} is a pair $(X_\bullet , Y_\bullet)$ where $X_\bullet$ is a Whitney stratification of $X$, $Y_\bullet$ is a  Whitney stratification of $Y$ and for each strata $M$ of $X$ there is a stratum $N$ of $Y$ with $f(M)\subset N$ such that the map $f|_M:M\to N$ is a {\em submersion}, i.e.~the differential $df|_M$ is surjective. It follows from the existence of the minimal Whitney stratification of a variety and, e.g.~the algorithm of \cite[\S 2]{helmer2023effective}, that there exists a unique minimal stratification of a map as well.

Given the defining equations of varieties $X$, $Y$ and of a map $f$ between them, the stratification $(X_\bullet, Y_\bullet)$ may be obtained explicitly using the algorithm of \cite{Helmer2021conormal,helmer2023effective} as implemented in the \texttt{WhitneyStratifications} Macaulay2 package.  Let us apply these to our toy example \eqref{eq:planarCubic}, where $X=\bV(f_z(x_1,x_2))\subset \RR^3$ and the map $f:X \to Y=\RR$ is given by the projection $(x_1,x_2,z)\mapsto z$. We find that the stratification of $Y=\RR$ is given by $\bV(z)\subset Y$,
hence in this example it correctly captures the change in topology at $z=0$. More generally, this is guaranteed thanks to {\em Thom's First Isotopy Lemma} \cite{mather2012notes}: If the map $f$ is proper (i.e if the inverse image of any compact subset is compact) then  the topology of the inverse image or fiber $f^{-1}(q)$ is constant over {\bf all} points $q\in N$ for any strata $N$ of $Y$ %\footnote{Note that in this example of \eqref{eq:planarCubic} the variety $X$ is affine, and hence its projection map to $\RR$ is not proper.   So as to apply Thom's isotopy lemma, we instead have to stratify $\overline{X}\to \CC$, where  $\overline{X}=\bV(x^{2}z\,w-x^{3}+y^{2}w-2\,y\,w^{2}+w^{3})$ is the projective closure of $X$ in $\PP^2\times \CC$, with $w,x,y$ coordinates on $\PP^2$.}
.

%%%%%%%%%%%%%%%%%%%%%%%%%%%%%%%%%%%%%%%%%%%%%%%%%%%%%%%%%%%%%%%%%%%%%%%%%%%%%%%%
%\section{\label{sec: Landau varieties}Whitney Stratification of Maps and the Landau Variety} 
Building on our earlier definitions and conventions, we now consider a Feynman integral specified by the homogenized Lee-Pomeransky polynomial $\G_h$ in the ring $\CC[z][x]:=\CC[z_1,\dots, z_m][x_0, \dots,x_n]$,  where the $z_i$ are external parameters (e.g.~masses, momenta) and the $x_i$ are the projective integration variables. 
%For a fixed vector of constants $z$ the resulting polynomial $\G_h$ thus defines a (projective) variety in the complex projective space $\PP^n$. 
We make no assumptions on the independence of the parameters, and in particular allow there to be algebraic relations between them. We then seek to describe the {\em Landau variety}, which is the locus in the parameter space $\CC_z^m$ where the Feynman integral is singular; the following definition is (a minor rephrasing of) that of Pham \cite[\S IV.5]{Pham2011integrals} for Feynman integrals in Lee-Pomeransky form:
\begin{definition}[Landau variety]\label{def: landau variety}
Consider a Feynman integral in homogenized Lee-Pomeransky form~\eqref{eq: projective Feynman integral}. Set $X=\bV(x_0\cdots x_n \G_h) \subset \PP_x^n\times \CC_z^m$ and consider the projection map $\pi:X \to \CC_z^m$ to the space of kinematics. The Landau variety is the variety $Y_{m-1}$ appearing in the minimal Whitney stratification $(X_\bullet, Y_\bullet)$ of the map $\pi$.    
\end{definition}

The virtue of the definition is that it renders the calculation of the Landau variety fully algorithmic. The relevant algorithms have been implemented in the \texttt{WhitneyStratifications} package, and include both obtaining the Whitney stratification of varieties and maps, and coarsening a Whitney stratification of a variety to the minimal one~\cite{helmer2024new}. 

It is also interesting to note that the lower dimensional strata in the stratification of Definition~\ref{def: landau variety} also have a physical significance: As we will see in an example shortly, they contain the Landau variety of integrals obtained as kinematic limits of the original one, e.g.~by setting some of its masses or momenta to zero. 
%That is, if $Y_{m-1}$ is the Landau variety of the original integral, then $Y_{m-2}$ contains the Landau variety of the integral obtained by setting some of its masses or momenta to zero, etc.
More generally, the stratification in Definition \ref{def: landau variety}	 describes all regions of the kinematic space such that any choice of parameters $z$ in a given region yields a variety $\pi^{-1}(z)$ in $\PP_x^n\times \CC_z^m$ with constant topology%\footnote{$\pi^{-1}(z)$ is the pairing of $z$ with a variety in $\PP_x^n$ corresponding to the specific $z$ value, thus we may think about it as a variety in $\PP_x^n$ parameterized by $z$}
.  This follows from the fact that the map $\pi$ in Definition \ref{def: landau variety} is proper by construction, as the fibers are (always compact) projective varieties and the aforementioned Thom's isotopy lemma always holds.

This observation has further implications when coupled with the result of \cite{Bitoun:2017nre}, that 
%(for parameter values outside the Landau variety) 
the Euler characteristic $|n+1-\chi(\bV(x_0x_1\cdots x_n\cG_h))|$ counts the number of basis or
master integrals~\cite{Chetyrkin:1981qh}
%, namely of linearly independent integrals with fixed propagator powers $\nu_i$
in the integral family~\eqref{eq: projective Feynman integral}.
Given the relation of lower strata with limits of the original integral, from their Euler characteristic computation we may obtain the number of master integrals also in these limits.

%%%%%%%%%%%%%%%%%%%%%%%%%%%%%%%%%%%%%%%%%%%%%%%%%%%%%%%%%%%%%%%%%%%%%%%%%%%%%%%%%%%%%%%%
\section*{Example computations}

We now demonstrate how the Landau singularities of several Feynman integrals may be calculated from Whitney stratifications of maps via  Definition \ref{def: landau variety}, and compare our results with existing approaches in the literature. More details on how these were carried out with the help of the \texttt{WhitneyStratifications} package may be found in the Supplemental Material. 

{\emph{One-loop bubble}}. We begin with this simple example,
\begin{center}
\begin{tikzpicture}[baseline=-\the\dimexpr\fontdimen22\textfont2\relax,transform shape,scale=0.95]
    \begin{feynman}
    \vertex (a);
    \vertex [right = of a] (b);
    \vertex [right = of b] (c);
    \vertex [right = of c] (d);
    \diagram* {
	    (a) --[fermion, edge label=\(p\)] (b) -- [half left,edge label=\({m_1}\)](c) -- [anti fermion, edge label=\(-p\)](d), (c) -- [half left,edge label=\({m_2}\)](b),
};
    \end{feynman}
    \end{tikzpicture}
\end{center}
illustrating that the Whitney stratification captures this integral's singularities not only for generic $m_1,m_2,p^2$, but also in the configurations where any of these kinematic variables vanish. Here, we have that $\cG_h=x_0(x_1+x_2)+(m_1^2+m_2^2-p^2)x_1x_2+m_1^2x_1^2+m_2^2x_2^2$,  $X=\bV(x_0x_1 x_2 \G_h) \subset \PP^2\times \CC^3$ and $Y=\CC^3$ is the space of kinematics. Calculating the (minimal) Whitney stratification $(X_\bullet , Y_\bullet)$ of the  projection map $\pi:X\to  \CC^3 $ yields the following the strata of $Y_\bullet$ below $Y_3=Y$, \small
\begin{align}\label{eq:bubble_stratification}
 Y_2=&\bV(m_1^2)\cup\bV(m_2^2)\cup\bV(p^2)\nonumber\\
    &\cup\bV(p^4+m_1^4+m_2^4-2p^2m_1^2-2p^2m_2^2-2m_1^2m_2^2),\nonumber\\
    Y_1=&\bV(p^2,\,m_1^2-m_2^2)\cup \bV(m_2^2-p^2,\,m_1^2)\cup \bV(m_2^2,\,m_1^2-p^2)\nonumber\\
    &\cup \bV(p^2,\,m_1^2)\cup \bV(p^2,\,m_2^2)\cup\bV(m_2^2,\,m_1^2),\nonumber \\
    Y_0=&\bV(p^2,\,m_1^2,\,m_2^2).
\end{align}
\normalsize
Per Definition \ref{def: landau variety}, the codimension-one components, $Y_2$, constitute the Landau variety. As already mentioned, the lower dimensional strata contain the Landau variety at certain limits. In the limit $m_1\to 0$, the Landau variety is read off from the three components in $Y_1$ containing a single $m_1^2$, and is particular equal to $\bV(m_2^2-p^2)\cup \bV(p^2)\cup \bV(m_2^2)$. In this example the result is  the same as substituting $m_1^2=0$ in the original Landau variety, but unlike the stratification the latter is not always guaranteed to work. Multiple limits can be treated in a similar fashion, e.g. components containing both  $m_1^2, m_2^2$ yield the Landau variety of the bubble with massless propagators, $\bV(p^2)$.

Our results are in complete agreement with the known Landau singularities of the generic one-loop bubble graph and its kinematic limits, see e.g.~\cite{Dlapa:2023cvx}, sections 4.1, 3.4  as well as the accompanying ancillary file. As also explained in the latter reference, Landau singularities precisely coincide with the rational letters of polylogarithmic integrals, whereas square-root letters are obtained by re-factorising products thereof. The precise re-factorisation rule is only known at one loop, with current attempts for a higher-loop generalisation known to be incomplete.

The Landau variety gives information of the full analytic structure of the meromorphic function defined by the integral.  However the methods in \cite{helmer2023effective} are capable of calculating Whitney stratifications of real semi-algebraic sets, potentially allowing direct access to singularities specifically in the physical region; we leave this fascinating exploration to a future work.

%%%%%%%%%%%%%%%%%%%%%%%%%%%%%%%%%%%%%%%%%%%%%%%%%%%%%%%%%%%%%%%%%%%%%%%%%%%%%%%%%%%%%%%%%%
{\emph{Slashed box.}}
We consider this first two-loop example, 
\begin{center}
\begin{tikzpicture}[baseline=-\the\dimexpr\fontdimen22\textfont2\relax]
            \begin{feynman}
            \vertex (v1) at ( {0.8*\xs}, {0.8*\xs});
            \vertex (v2) at (-{0.8*\xs}, {0.8*\xs});
            \vertex (v3) at (-{0.8*\xs},-{0.8*\xs});
            \vertex (v4) at ( {0.8*\xs},-{0.8*\xs});

            \vertex (i1) at ( {1.6*\xs}, {1.6*\xs}){\(p_4\)};
            \vertex (i2) at (-{1.6*\xs}, {1.6*\xs}){\(p_1\)};
            \vertex (i3) at (-{1.6*\xs},-{1.6*\xs}){\(p_2\)};
            \vertex (i4) at ( {1.6*\xs},-{1.6*\xs}){\(p_3\)};
    \diagram*{
        (v1)--(v2)--(v3)--(v4)--(v1), (v1)--(v3),
        (i1) -- [fermion](v1), (i2) -- [fermion](v2),(i3) -- [fermion](v3), (i4) -- [fermion]  (v4);
    };
    \end{feynman}
    \end{tikzpicture}
\end{center}
in the ``two-mass hard'' configuration with $p_3^2\neq0,\ p_4^2\neq 0$ and everything else massless. In this example, $X=\bV(x_0\cdots x_5 \G_h) \subset \PP^5\times \CC^4$, with $Y=\CC^4$ spanned by $p_3^2, p_4^2, s=(p_1+p_2)^2, t=(p_2+p_3)^2$. The Whitney stratification $(X_\bullet , Y_\bullet)$ of the projection map $\pi:X\to  \CC^4 $ gives the following expression for the Landau variety,
\small
\begin{align}
    Y_3=&\bV (p_3^2)\cup \bV( s)\cup \bV(st+t^2-tp_3^2-tp_4^2+p_3^2p_4^2)\nonumber\\&\cup  \bV(t-p_4^2) \cup \bV(s^2-2sp_3^2+p_3^4-2sp_4^2-2p_3^2p_4^2+p_4^4)\nonumber\\ &\cup \bV(t-p_3^2)\cup \bV(t)\cup \bV(p_4^2)\cup \bV(p_4^2-s-t)\label{eq:SlashedBoxY}.
\end{align}\normalsize
This is in agreement with \texttt{HyperInt}, whereas \texttt{PLD} misses the component $p_4^2-s-t=0$. That this is a true singularity of the integral may be inferred e.g.~by its explicit analytic expression~\cite{Henn:2014lfa}; therefore this nontrivial example strongly supports the completeness of our approach for computing Landau singularities.

We note that the empirical procedure proposed in \cite{Dlapa:2023cvx} for the calculation of Landau singularities also reproduces~\eqref{eq:SlashedBoxY}. 
%This procedure starts with the principal $A$-determinant of the polynomial $\cG$ in \eqref{eq: Feynman integral}, generalised such that all of its coefficients become distinct indeterminates, and applies sequential limits to set them to the values of the corresponding Feynman integral.
%We observe that we arrive at the same conclusion also when starting from the product of Landau singularities of the generic slashed box with all internal and external masses nonzero and different from each other. 
It would be very interesting to investigate if its coincidence with our rigorous Definition~\ref{def: landau variety} holds more generally.

%%%%%%%%%%%%%%%%%%%%%%%%%%%%%%%%%%%%%%%%%%%%%%%%%%%%%%%%%%%%%%%%%%%%%%%%
{\emph{Parachute integral.}}  This is another two-loop example,
\begin{center}
\begin{tikzpicture}[baseline=-\the\dimexpr\fontdimen22\textfont2\relax]
            \begin{feynman}
            \vertex (v1) at ( -\xs, 0);
            \vertex (v2) at ( {.5*\xs}, {-0.86602540378*\xs});
            \vertex (v3) at ( {.5*\xs}, {0.86602540378*\xs});

            \vertex (i1) at ( {-1.9*\xs}, 0){\(p_3\)};
            \vertex (i2) at ( {1.9*\xs/2}, {-1.9*\xs*0.86602540378}){\(p_2\)};
            \vertex (i3) at ( {1.9*\xs/2}, { 1.9*\xs*0.86602540378}){\(p_1\)};
    \diagram*{
        (v1)[dot]--[edge label'=\(m_4\)](v2)--[edge label'=\(m_1\)](v3)--[edge label'=\(m_3\)](v1),(v2)--[half right, edge label'=\(m_2\)](v3),
        (i1) -- [fermion](v1), (i2) -- [fermion](v2),(i3) -- [fermion](v3);
    };
    \end{feynman}
    \end{tikzpicture}
\end{center} 
whose singularities are not entirely captured by either the naive principal $A$-determinant~\cite{Klausen:2023gui}, or its principal Landau determinant refinement. In particular, these approaches miss the singularity where
{\small
\begin{equation}\label{eq: parachute discriminant}
    p_3^2(m_4^2 - p_2^2)(m_3^2 - p_1^2) + (m_3^2 - m_4^2 - p_1^2 + p_2^2)(m_3^2p_2^2 - m_4^2p_1^2)
\end{equation}}vanishes, first computed in \cite{Berghoff:2022mqu}. The authors of this reference define the Landau variety by demanding that the union of the (integration) coordinate hyperplanes $x_i=0$ with the hypersurfaces $\cU=0$ and $\cF=0$ is \emph{transverse}, otherwise \emph{blowing up} this manifold in order to ensure this property; it turns out that the parachute requires additional blow-ups beyond what was previously considered sufficient for Feynman integrals. We note that the method of blow-ups is an equivalent alternative method to calculate the Whitney stratification in our context, since in the transversal case the Whitney stratification has a simple closed form expression%~\footnote{As additional evidence for the presence of the singularity~\eqref{eq: parachute discriminant}, we also mention that it is an output of  \texttt{HyperInt}, and we have checked that the Euler characteristic of the hypersurface complement drops from the generic value 19 to the value 18 on it.}
.

The stratification of the map $\pi:\ \PP^4\times\CC^7\to \CC^7$ is currently too taxing for the \texttt{WhitneyStratifications} package, however we will validate our approach by restricting the kinematics to  $\{m_1^2=1,\,m_2^2=0,\,m_3^2=0,\,m_4^2=2,\,p_1^2=-1,\,p_2^2=0\}$, keeping only $p_3^2$ free. In this limit, the singularity \eqref{eq: parachute discriminant} reduces to $p_3^2-1$, and direct integration  confirms its presence as a letter of the integral \footnote{We thank Erik Panzer for confirming this.}. We confirm that this is also contained in our Whitney stratification,
\begin{equation*}
Y_0=\bV(p_3^2)\cup\bV(p_3^2+2)\cup\bV(p_3^2-2)\cup\bV(p_3^2+1)\cup\bV(p_3^2-1)\,;
\end{equation*}
for comparison, \texttt{PLD} appears to predict only the first three of these singularities.

%%%%%%%%%%%%%%%%%%%%%%%%%%%%%%%%%%%%%%%%%%%%%%%%%%%%%%%%%%%%%%%%%%%%%%%%
\section*{ Conclusions and Outlook}
We presented an algorithm to calculate the Landau variety of an integral, from the Whitney stratification of the projection map on its integrand. Applicable to any parameterized integral with a rational integrand, this provides the complete and non-redundant set of Landau singularities of not only the original integral, but also of those arising as limits of its parameters. 

Leveraging the software implementation of Whitney stratifications, we focused on Feynman integrals and validated our method in several examples through two loops, including cases where singularities are missed by other approaches. It would be exciting to improve our method's efficiency by taking advantage of the special structure of Feynman integrals. Given the relation between Landau singularities and symbol letters established in~\cite{Dlapa:2023cvx}, could it further be used to predict the latter, and prove the remarkable cluster-algebraic structures they sometimes exhibit~\cite{Chicherin:2020umh}?

%%%%%%%%%%%%%%%%%%%%%%%%%%%%%%%%%%%%%%%%%%%%%%%%%%%%%%%%%%%%%%%%%%%%%%%%
\begin{acknowledgments}
MH and FT would like to thank the Isaac Newton Institute for Mathematical Sciences, Cambridge, for support and hospitality during the program ``New equivariant methods in algebraic and differential geometry" where work on this paper was undertaken. This visit was supported by EPSRC grant no EP/R014604/1. 

MH was supported by the Air Force Office of Scientific Research (AFOSR) under award
number FA9550-22-1-0462, managed by Dr.~Frederick Leve, and would like to gratefully acknowledge this support. 

The authors would also like to thank Rigers Aliaj, Saiei Matsubara-Heo, Erik Panzer and Simon Telen for helpful discussions during the preparation of this letter.
\end{acknowledgments}

\bibliography{Ref.bib}% Produces the bibliography via BibTeX.

\onecolumngrid

\section*{Supplemental material}

For completeness, here we provide specialized information on the application of the \texttt{WhitneyStratifications} Macaulay2~\cite{M2} package to compute the Landau singularities of Feynman integrals. In the discussion below we use the most recent version of this package, version 2.27, which is available at: \url{http://martin-helmer.com/Software/WhitStrat/WhitneyStratifications.m2}. We will demonstrate the relevant commands on the example of the generic bubble integral,  also discussed in the main text.

First, we load the package,
\begin{lstlisting}
needsPackage "WhitneyStratifications"
\end{lstlisting}
Then, we define the variety $X:=\texttt{X}$ in the space of both external kinematic parameters and integration variables, the variety $Y=\bV(0)\subset\CC^3=\CC^3:=\texttt{Y}$ as the space of external kinematic parameters, as well as the projection map $\pi:=\texttt{piMap}$ from $X$ to $Y$, in terms of the relevant rings and ideals:
\begin{lstlisting}
R = QQ[M1,M2,P,x_0..x_2]
U = (x_1+x_2)
F = -P*x_1*x_2+U*(M1*x_1+M2*x_2)
Gh = U*x_0+F
X = ideal(x_0*x_1*x_2*Gh)
params = {M1,M2,P}
S = QQ[params]
piMap=params
Y = ideal(0_S)
\end{lstlisting}
For simplicity we use squared momenta and masses as parameters, $\texttt{P}=p^2$, $\texttt{M1}=m_1^2$ etc. Also note that we need to manually define the variables of the rings $R$ and $S$ even though those of the latter are contained in those of the former, so as to avoid conflicts with variable naming conventions in Macaulay2.

The relevant function of the package for the Whitney stratification of maps is \texttt{mapStratify}, for documentation see \url{http://martin-helmer.com/Software/WhitStrat/_map__Stratify.html}\label{foot:mapStratify}. The default command, implementing the algorithm~\cite{Helmer2021conormal} for the calculation of map stratifications, in our case reads:
\begin{lstlisting}
ms=mapStratify(piMap,X,Y)
\end{lstlisting}

Owing to its completeness and applicability to any variety and map, this command is in general quite time consuming~\footnote{That is to say that the general algorithm on the one hand does not yet take advantage of any special features enjoyed by the varieties associated to Feynman integrals, and on the other hand computes all strata, not just the one corresponding to the Landau variety.}. Very interestingly, for all examples of Feynman integrals considered, we have observed that it is enough to calculate the stratification of the corresponding map imposing only that the successive differences of elements in each of the flags $X_\bullet, Y_\bullet$ are smooth, i.e. by computing consecutive singular loci. While there is currently no theoretical guarantee for its completeness, this version of the stratification is significantly faster, and may also be found as an option of the \texttt{mapStratify} function:
\begin{lstlisting}
ms=mapStratify(piMap,X,Y,StratsToFind=>"singularOnly")
\end{lstlisting}

\iffalse
This faster, but potentially incomplete, version of the stratification can be found as an option in the  \texttt{mapStratify} function. In particular, to speed up the computation we have several options. The first is to use the \texttt{StratsToFind} option, which takes inputs \texttt{"all"}, \texttt{"most"}, and \texttt{"singularOnly"}.
The last option is the fastest in general, but may miss strata. The option \texttt{"most"} is very likely to return all strata (but it is not guaranteed), while the option \texttt{"all"} is guaranteed to find all strata but the stratification may fail to be minimal and the run times may be extremely long. For this example the \texttt{"singularOnly"} option below returns the correct and minimal stratification of $\pi$ in less than 10 seconds (on a laptop):
\begin{lstlisting}
ms=mapStratify(piMap,X,Y,StratsToFind=>"singularOnly")
\end{lstlisting}
If we instead run 
\begin{lstlisting}
ms=mapStratify(params,X,piMap,StratsToFind=>"most")
\end{lstlisting}
we will again get the correct and minimal stratification of $\pi$ in less than 10 seconds. 
\fi

Yet a third possibility that is in the middle ground with respect to speed, but is guaranteed to find a complete and minimal stratification, relies on the recent variation of the main algorithm presented in \cite[\S3]{helmer2023effective}. This is implemented in the \texttt{mapStratifyPol} command, which is only present in the most recent version of the \texttt{WhitneyStratifications} package. More precisely, a single run of the algorithm \cite[\S3]{helmer2023effective} produces a complete, non-minimal stratification where the unnecessary strata generated are random. Hence the minimal stratification may be obtained by running the command twice and taking the common elements of both computations, which in our example may be accomplished as follows: 
\begin{lstlisting}
ms1=mapStratifyPol(piMap,X,Y)
ms2=mapStratifyPol(piMap,X,Y)
ms={new MutableHashTable from 
for k in keys first ms1 list k=>toList(
(set(first ms1)#k)*(set(first ms2)#k)),
new MutableHashTable from
for k in keys last ms1 list k=>toList(
(set(last ms1)#k)*(set(last ms2)#k))}
\end{lstlisting}
One run of the \texttt{mapStratifyPol} command on this example (on a laptop with an Intel i7-1185G7 processor and 32GB of RAM) takes 212 seconds, in other words the block of code above takes around 424 seconds to execute.

Irrespective of which of the three possible methods described above is used, we store the resulting stratification
 $(X_\bullet, Y_\bullet)$  in the variable \texttt{ms}, which is a list containing two objects of type \texttt{MutableHashTable}, with \texttt{ms\textunderscore0} representing $X_\bullet$ and \texttt{ms\textunderscore1} representing $Y_\bullet$. To view the contents of $Y_\bullet$ we run:
\begin{lstlisting}
peek ms_1
\end{lstlisting}
which returns: 
\begin{lstlisting}
MutableHashTable{0 => {ideal(P,M2,M1)}                                                
    1 => {ideal(P,M1-M2),ideal(M2-P,M1),ideal(M2,M1-P),ideal(P,M1),ideal(P,M2),ideal(M2,M1)}
                                               2         2             2
    2 => {ideal P, ideal M1, ideal M2, ideal(M1-2M1*M2+M2-2M1*P-2M2*P+P)}
    3 => {ideal 0}}
\end{lstlisting}
The integer index in each element of the output table above amounts to the dimension of the corresponding component of $Y_\bullet$. Therefore the aforementioned output is in perfect agreement with the result for the latter quoted in eq. (5) of the main article.

We observe that the default algorithm~\cite{Helmer2021conormal} also yields a minimal stratification in this example. In fact, we expect that this will be tha case for any Feynman integral, and while we have not proven this, we have checked its correctness in all cases considered. In contradistinction, we have also established that the alternative Whitney stratification algorithm described in~\cite{helmer2024new}, while complete and correct, does yield non-minimal straficications for certain Feynman integrals. In our example, this alternative algorithm could be run as follows:
\begin{lstlisting}
ms=mapStratify(piMap,X,Y,AssocPrimes=>false)
\end{lstlisting}

We wish to comment that non-minimality is not an obstacle in principle, as an algorithm to minimize any given Whitney stratification of a variety is given in \cite[Section~3.3]{helmer2024new} and is implemented in the \texttt{minCoarsenWS} command of the \texttt{WhitneyStratifications} package. Nevertheless, care is needed when using this command so as to minimize Whitney stratifications of maps: Instead of naively applying it to each of the two outputs of the map stratification, one needs to keep track of which of the strata are intrinsically associated to the varieties involved, and which are associated to critical values of the projection map between them. The \texttt{minCoarsenWS} command should only be applied to the former, and a workaround to the fact that no information on the origin of the strata is currently recorded, is to do a separate variety stratification calculation (this should be contained in the map stratification).  Ideally, the test \cite{helmer2024new} should be performed ``live" during the map stratification computation, i.e.~checking each newly computed Whitney stratum to see if it is actually required as the process runs. A fully automated and efficient function to minimize the stratification of a map, which works in this manner, is planned to be implemented in future versions of the package. 

\end{document}